\documentclass[12pt]{iopart}

\usepackage{iopams}  
\usepackage{silence}
\usepackage{natbib}
\expandafter\let\csname equation*\endcsname\relax
\usepackage{graphics}
\usepackage{graphicx}
\usepackage{subfigure}
\usepackage{color}
\usepackage{soul}
\setstcolor{red}
\usepackage{comment}
\bibpunct{(}{)}{;}{a}{,}{,}
\usepackage{hyperref}
\hypersetup{
    colorlinks=true,
    linkcolor=blue,
    filecolor=magenta,      
    urlcolor=cyan,
}
\usepackage{array}
\newcolumntype{C}[1]{>{\centering\arraybackslash}p{#1}}
\usepackage{graphicx}
\usepackage{algorithm}

\usepackage{multirow}
\makeatletter
\newcommand{\quickwordcount}[1]{%
  \immediate\write18{texcount -1 -sum -merge #1.tex > #1-words}%
  \immediate\openin\somefile=#1-words%
  \read\somefile to \@@localdummy%
  \immediate\closein\somefile%
  \setcounter{wordcounter}{\@@localdummy}%
  \@@localdummy%
}

\usepackage{makecell}
\definecolor{Silver}{rgb}{0.9,0.9,0.9}
\definecolor{Gray}{rgb}{0.75,0.75,0.75}
\definecolor{LightCyan}{rgb}{0.8,1,1}
\definecolor{LightGreen}{rgb}{0.8, 1, 0.8}

\providecommand{\add}[1]{{\color{black}#1}}

\makeatother

\begin{document}

\font\myfont=cmr12 at 16pt

\title[Open-Source Activity Classification from Triaxial Accelerometry]{An Open-Source, Open Data Approach to Activity Classification from Triaxial Accelerometry in an Ambulatory Setting}


\author{Sepideh Nikookar$^{1}$,
Edward Tian$^{2}$,
Harrison Hoffman$^{2}$, Matthew Parks$^{2}$, J. Lucas McKay$^{1,3}$, Yashar Kiarashi$^{1}$,\\
Tommy T. Thomas$^{1,4}$,
Alex Hall$^{5}$, 
David W. Wright$^{5}$\\
and Gari D. Clifford$^{1,6}$\\}

\address{$^{1}$Department of Biomedical Informatics, Emory University, Atlanta, GA, USA} 
\address{$^{2}$LifeBell AI LLC, Atlanta, GA, USA}
\address{$^{3}$Department of Neurology, Emory University, Atlanta, GA, USA} 
\address{$^{4}$Department of Neurosurgery, Emory University, Atlanta, GA, USA}
\address{$^{5}$Department of Emergency Medicine, Emory University, Atlanta, GA, USA}
\address{$^{6}$Department of Biomedical Engineering, Georgia Institute of Technology, Atlanta, GA, USA}

\ead{gari@gatech.edu}

\vspace{10pt}


\begin{abstract}

The accelerometer has become an almost ubiquitous device, providing enormous opportunities in healthcare monitoring beyond step counting or other average energy estimates in 15-60 second epochs. 

\add{
\underline{Objective}: To develop an open data set with associated open-source code for processing 50 Hz tri-axial accelerometry-based to classify patient activity levels and natural types of movement.}

\add{
\underline{Approach}:}
Data were collected from 23 healthy subjects (16 males and seven females) aged between 23 and 62 years using an ambulatory device, which included a triaxial accelerometer and synchronous lead II equivalent ECG for an average of 26 minutes each. Participants followed a standardized activity routine involving five distinct activities: lying, sitting, standing, walking, and jogging. Two classifiers were constructed: a signal processing technique to distinguish between high and low activity levels and a convolutional neural network (CNN)-based approach to classify each of the five activities. 

\add{
\underline{Main results}:}
The binary (high/low) activity classifier exhibited an F1 score of 0.79. The multi-class CNN-based classifier provided an F1 score of 0.83. The code for this analysis has been made available under an open-source license together with the data on which the classifiers were trained and tested.

\add{
\underline{Significance}:}
The classification of behavioral activity, as demonstrated in this study, offers valuable context for interpreting traditional health metrics and may provide contextual information to support the future development of clinical decision-making tools for patient monitoring, predictive analytics, and personalized health interventions.

\end{abstract}

%
\vspace{2pc}

\maketitle


\section{Introduction}
Continuous monitoring of physiological parameters plays a vital role in modern healthcare, enabling earlier detection of clinical deterioration and supporting personalized treatment strategies. Traditional biomedical monitoring approaches often relied on sporadic measurements collected during clinical visits or hospitalizations. However, advances in sensing technologies have increasingly enabled continuous, real-time data collection \cite{grana2020use, goldsack2020verification, teixeira2021wearable}, potentially transforming both clinical and home-based patient care.

These technological innovations span two major domains: on-body and off-body sensors. On-body wearable devices, such as smartwatches~\cite{jat2022smart} and in-ear EEG systems~\cite{joyner2024using,waters2024domain}, capture physiological signals directly from the body. Off-body sensors, including thermal cameras~\cite{manullang2021implementation} and radar systems~\cite{fraccaro2020development}, offer non-contact methods for monitoring patients from a distance. Alternative non-contact modalities, such as vision-based \cite{sathyanarayana2018vision}, audio-based \cite{carron2021mobile}, and radar/LiDAR-based systems \cite{Li_2013_6504804,Dong_2024_10531059,Rinchi_2023_10145011}, have been explored to detect activity patterns and vital signs without requiring physical contact.

Despite these advancements, non-contact monitoring faces critical limitations in real-world healthcare environments. Vision-based systems are sensitive to occlusions and lighting variability, while radar- and LiDAR-based methods, though promising, often suffer from environmental clutter and privacy concerns. These challenges can limit the reliability of non-contact approaches, particularly in dynamic and crowded clinical settings.

As a result, wearable sensors remain the most practical and scalable solution for continuous patient monitoring in many environments. Accelerometers, in particular, have garnered significant attention for their ability to provide real-time insights into patient activity and movement patterns. Accelerometers are widely used in remote patient monitoring, and to some extent in emergency department waiting rooms \cite{10.1109/iembs.2010.5627285,10.3389/fphys.2015.00149, curtis2008smart, nino2020coupling}, and home-based care settings \cite{albahri2019based,wang2017review}. By capturing continuous data on patient motion, accelerometers offer valuable context for interpreting physiological parameters such as heart rate and oxygen saturation \cite{ivașcu2021activity}. For example, knowing whether a patient is walking, standing, or lying down can help clinicians accurately attribute changes in heart rate, reducing the risk of false alarms or misinterpretations.

However, even wearable sensors are not immune to challenges. Motion artifacts caused by natural patient movements can introduce noise into telemetry data, complicating real-time assessments \cite{freeman2006autonomic,pawar2007impact}. Movements such as walking, shifting posture, or adjusting clothing can interfere with the accurate monitoring of heart rate, respiratory rate, and oxygen saturation. Although studies have validated the reliability of wearable devices compared to traditional clinical monitors, demonstrating comparable heart rate and respiratory rate measurements in intensive care unit settings \cite{stevens2024feasibility}, real-world wearable data often suffer from noise, missingness, and imbalanced class distributions \cite{taffoni2018wearable}. These data quality issues can degrade model performance, particularly in ambulatory health monitoring applications where environmental variability is high.

Human activity recognition (HAR) plays a crucial role in mitigating these challenges by providing context for interpreting physiological data. However, existing HAR methods face several limitations. Many models struggle with small labeled datasets, difficulty capturing temporal dependencies, computational constraints for real-time deployment, and the impact of noise on recognition accuracy \cite{arshad2022human}. While architectures such as LSTM-based models have shown promise \cite{mekruksavanich2021lstm,ReyesOrtiz2013}, their effectiveness often relies on high-quality, synchronized time-series data, which may not be available in real-world ambulatory settings. Similarly, multi-sensor fusion strategies—such as combining accelerometry with ECG data—have improved classification performance in specific contexts \cite{ren2024clinical}, but require complex hardware configurations that may not be feasible in resource-limited environments.

To address these limitations, this study focuses on developing robust, scalable activity recognition methods based solely on 3D accelerometry data. CNN-based approaches have demonstrated strong potential for human activity recognition across a variety of domains, capturing temporal and spatial patterns in sequential sensor data \cite{shi2023novel, ismail2023auto, bao2004activity, yuan2022interpretable, zhu2018novel}. These strategies aim to balance classification accuracy, computational efficiency, and practical feasibility for real-world deployment in ambulatory health monitoring systems.

While prior studies, such as \cite{ren2024clinical}, demonstrated strong performance using accelerometer-based models, their datasets \cite{Ren2023} were collected in controlled laboratory environments with balanced data distributions and longer per-participant recordings. In contrast, this work targets the challenges of noisy, imbalanced data collected from shorter, more variable ambulatory recordings, which more closely reflect the data characteristics expected in real-world health monitoring systems. Rather than attempting to outperform models trained on high-fidelity, controlled datasets, this study focuses on developing robust methods that generalize under the imperfect and unpredictable conditions that commonly arise in practical healthcare deployments.

\add{Recent work has explored inertial sensing for fine-grained motion and gesture recognition using wearable systems, including approaches that combine accelerometry with additional physiological signals \cite{chen2025noise}. However, these studies primarily focus on discrete gesture recognition or structured activity sequences. In contrast, the present work emphasizes sustained activity states in noisy, real-world ambulatory settings. The goal is not gesture recognition or multimodal sensor fusion, but the development of robust accelerometry-only methods that support contextual interpretation of physiological data in practical healthcare environments.}

Similarly, \cite{luckhurst2024classifying} developed accelerometer-based cut-points for binary activity classification using a Mean Amplitude Deviation (MAD) metric. In contrast, this study applies a rolling 3D magnitude method that can be implemented in near real-time, producing classification windows with a fixed delay of (N - 1)/2 samples. For the 5 s window at 50 Hz (N = 250 samples), this corresponds to a delay of approximately 2.5 s, which could be considered negligible for real-time applications. While MAD captures variability around the mean acceleration, the 3D magnitude reflects overall movement intensity and is easier to interpret in streaming applications. The threshold used in this study was determined empirically by analyzing distributions from early pilot data collected before the main study. This value was then spot-checked against the study dataset to ensure it generalized well to the broader population.

The objective of this study was to develop and release open data alongside open-source accelerometry-based methods for accurately classifying activity levels and specific movement types encountered in routine ambulatory settings. By automatically creating activity-type labels, these methods may improve the contextual interpretation of physiological signals. Such contextualization can enhance clinical decision-making, reduce false alarms, and support more personalized health interventions in diverse healthcare environments.

To this end, two complementary approaches were implemented: a real-time, a signal processing–based method to distinguish between high and low activity levels through cut-point–based classification, and a CNN-based deep learning model for multi-class activity recognition. While the cut-point–based method is computationally efficient and suitable for real-time deployment, the deep learning model enables more granular and clinically meaningful behavioral assessments.

\add{To the best of our knowledge, this is the first study to release both open-access triaxial accelerometry data collected in a realistic ambulatory setting and fully reproducible open-source code for human activity classification. This dual release enables complete transparency, reproducibility, and independent benchmarking under noisy, imbalanced, and non-stationary conditions that more closely reflect real-world healthcare environments. Furthermore, the proposed methods were intentionally designed with computational efficiency in mind, supporting potential deployment on low-cost edge devices for real-time monitoring applications.}

\section{Methods}\label{sec:method}
\subsection{Device Details}

A practical and reusable wearable ECG monitor cardiac patch (the Vivalink VV330 Continuous ECG Platform and VivaLNK Adhesive Patch) was used for data collection. This patch serves as a versatile tool for gathering physiological data. It records single lead ECG data at a rate of 128 Hz, providing detailed insights into heart activity. Additionally, it captures triaxial accelerometry data at a rate of 50 Hz, which is the sampling rate provided by the device manufacturer, to offer valuable information about body movements and activity levels. Generally, human movement is confined to 20 Hz or less \cite{antonsson1985frequency}. A Nyquist frequency of 25 Hz is therefore sufficient to capture all relevant movement.

Furthermore, the patch automatically calculates heart rate measurements at a frequency of 1 Hz based on the recorded ECG signals. This feature enables real-time monitoring of heart rate variations, enhancing the depth of information collected during the study.

Although the device collects both ECG and accelerometry signals, only the accelerometry data were utilized for activity classification in this study. The ECG signals were analyzed separately to assess physiological parameters, such as heart rate deviations, to quantify the physiology of the population we used, but they were not fused with motion data for activity recognition. While both data streams were available from the same hardware unit, the computational and modeling simplicity of our approach stems from using only accelerometry for real-time activity classification. This separation ensures that the proposed activity recognition methods remain scalable to much cheaper devices that collect only accelerometry.

\subsection{Data Collection Protocol}

Participants were recruited through flyers distributed in the office of Prophecy Games, an online computer game company affiliated with one of the authors, and in neighboring offices within the same building. This company provided only the office space for flyer distribution and had no role in the study design, data collection, analysis, or interpretation; there is no financial or scientific conflict of interest related to this arrangement. These flyers were approved by the Institutional Review Board (IRB) and invited individuals to participate in the study. A total of 23 participants (16 males and seven females), aged between 23 and 62 years, were enrolled. Relevant physiology (age, weight, and height) were recorded.
 However, demographic variables such as race and education level were not collected, which limits the ability to assess model generalizability across different populations.
 
 The mean age of participants was $32.44\pm 9.84$ years for males and $36.00 \pm 12.01$ years for females.
  All participants provided informed consent prior to participation, and the study received ethical approval from WCG IRB (Pr. No.: 20231639). Each participant engaged in a structured series of activities to replicate common scenarios encountered in non-sedentary hospital environments, such as those in emergency departments. These activities included a range of movements and postures typical of such environments, ensuring a representative dataset for analysis.

Prior to data collection, subjects underwent a standard skin cleaning procedure with an alcohol swab to optimize contact between the ECG patch electrodes and the skin. Each subject was fitted with a single patch, securely applied to approximate the Lead II position on their left chest using a specially designed adhesive patch. This standardized application method ensured consistent electrode positioning across all subjects, which is crucial for reliable data collection. Since each participant’s data collection occurred independently, any malfunction or error with an individual device would only affect that participant’s data, while the data from other participants would remain unaffected. This study design ensured that the overall dataset remained robust despite potential issues with individual devices. To further enhance data quality, participants with excessive chest hair were shaved to ensure optimal contact between the patch and the skin; this was essential since excessive hair could disrupt data collection.

Before commencing the study, thorough checks were performed to confirm the smooth transmission of data from the patch to the recording system and temporal continuity of each data packet. This step was crucial to ensure the accuracy and consistency of the data, establishing a reliable basis for the subsequent analysis. Data were streamed to a smartphone via Bluetooth and offloaded to AWS for secure storage and analysis.

Subjects were guided through various activities in an office environment that mirrored the dynamic nature of patient behavior in emergency department waiting areas. These activities included lying down, sitting, standing, walking, and jogging. While subjects sat in office chairs and laid on a couch during the sedentary tasks, walking and jogging activities were specifically conducted in a larger conference room, where subjects walked back and forth or in circles to further simulate real-world conditions. Walking in circles was included as a controlled way to simulate movement in a confined space, which can be representative of certain real-world conditions, particularly in environments where space is limited, such as hospital waiting areas. This activity was intended to capture the physical exertion and movement patterns of individuals in dynamic, constrained spaces. While walking in circles may not universally represent all real-world scenarios, it provides a practical method to induce movement that mimics some aspects of real-world activity. Instructions were provided for participants to perform fidgeting behaviors, such as checking their mobile phones, bouncing their legs, retying their shoes, or adjusting their clothes or hair, to capture the subtleties of real-world scenarios.

To simulate potential disruptions, subjects were asked to tap on the patch or scratch the skin around it. This deliberate interference was designed to test the robustness of the monitoring system under challenging conditions, providing insights into its reliability and resilience in practical settings.

On average, each subject contributed $26.33 \pm 21.36$ minutes of continuous ECG and accelerometry data throughout the activity protocol. The minimum duration of data collected from an individual participant was 4.22 minutes, while the maximum duration was 67.98 minutes. The large standard deviation reflects the variation in participation, as some subjects were not comfortable completing all activities outlined in the protocol, resulting in shorter data collection duration. This data was further complemented by synchronized video recordings, which served as ground truth references for accurate labeling and validation. Videos of the subjects’ activities were recorded on an iPad, and the iPad and patch were aligned with the world clock before every session to ensure that the video and accelerometry data were as time-aligned as possible. The combination of video and physiological data facilitated a comprehensive analysis of motion-related artifacts, enabling a thorough investigation into how motion affects biomedical data interpretation.

 \add{It is important to note that the present dataset was collected from healthy volunteers performing structured activities in a simulated office environment. While the activities were designed to resemble ambulatory patient behaviors, the findings should not be interpreted as direct validation in clinical patient populations.}

\subsection{Labeling}

The activity data collected was carefully labeled by hand, with close reference to the ground truth videos, to ensure accuracy. To label the data, an open-source tool, Label Studio, was used, which allowed for collaboration among team members and provided transparency in the labeling process. While only two individuals were responsible for the actual labeling, the platform remained accessible to the entire research team, allowing for oversight and review by qualified personnel, including a registered nurse. The data was initially grouped into different activity states: lying, sitting, standing, walking, and jogging. Each of these states was considered separately, without any overlap between them.

\begin{figure}[!htbp]
\centering
\includegraphics[width=0.5\textwidth]{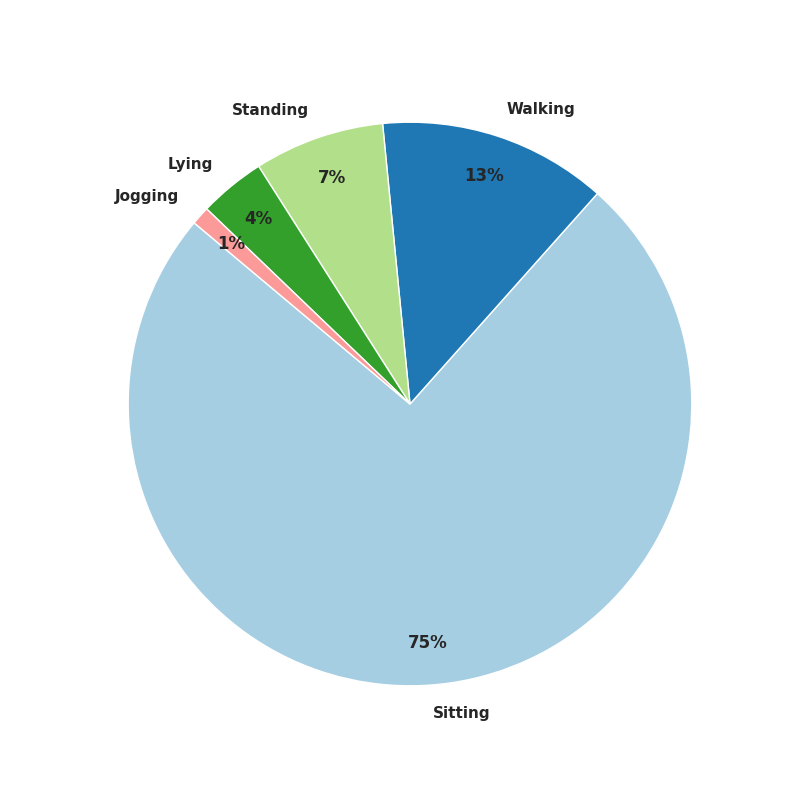}
\caption{Activity label distribution across the dataset.}
\label{fig:label_distribution}
\end{figure}

The distribution of these activity states across the dataset is shown in Figure \ref{fig:label_distribution}. This pie chart visually represents the proportions of time each subject spent in each activity during the data collection process, providing insights into the relative frequency of each activity.

In parallel, the data were categorized based on the subjects' posture. This included whether they were leaning forward, leaning backward, leaning to the left, or leaning to the right. If a specific posture was not identified, it was assumed the subjects were upright.

In addition to categorizing activity and posture, labels were assigned to capture general activities during the data collection process. The label assigned to time $t$ was whatever activity the subject performed in the window from $t - 5$ to $t$ seconds. These labels encompassed transitions between different postures, instances where the device might have been interfered with, and other actions like fidgeting or adjusting one's position. The 5-second window was selected to balance capturing sufficient temporal context for each activity while maintaining precision in detecting activity transitions. Larger window sizes could lead to diminishing returns by smoothing out subtle transitions, making classification less accurate. 

While the vast majority of windows only contained a single activity, as intended by the protocol, there were windows where the subject transitioned between activities. In such cases, the window was labeled according to the activity that occurred most recently within the window. If a clear transition between two activities was observed, the window was labeled as a transition. Transition-labeled windows were excluded from subsequent model training and evaluation to ensure that the models were trained on well-defined, stable activity states. 
\add{This exclusion could be seen as providing optimistic performance estimates compared to real-world continuous monitoring scenarios, where transitions between activities are frequent and unavoidable. However, as noted in the discussion, this is an open problem with all current classifiers.}

Importantly, the activity labels (lying, sitting, standing, walking, and jogging) were mutually exclusive, meaning that a window could only be labeled with one of these activity states at a time. While participants were instructed to perform natural fidgeting behaviors, these micro-movements were not assigned independent labels in the final dataset. Instead, windows containing fidgeting were labeled based on the primary activity the subject was engaged in at the time. This decision was made to ensure consistency in activity classification, as fidgeting often overlapped with multiple activity states and did not represent a distinct category within the model framework.


\subsection{Preprocessing}
\label{sec:preprocess}

To prepare the data for analysis, triaxial accelerometry signals collected from each subject were utilized. The preprocessing steps were crucial to ensure the data was clean and reliable for subsequent analysis.

First, all three axes of the accelerometry signals were preprocessed by a Butterworth bandpass filter with cutoff frequencies set at 0.05 Hz and 2 Hz. The low-frequency cutoff of 0.05 Hz removed sensor drift, while the high-frequency cutoff of 2 Hz filtered out irrelevant signals, such as high-frequency vibrations from devices like phones or nearby machinery. Studies have shown that most human motion-related signals, including walking, running, sitting, and standing, predominantly fall below 20 Hz, as reported in frequency domain analyses of human activity recognition \cite{antonsson1985frequency}. To further validate this choice, a sensitivity analysis was conducted by varying the high-frequency cutoff to 2, 5, 10, 15, and 20 Hz. The results of this analysis are presented in Section \ref{sec:cutoff}. This filtering approach preserved essential physiological or motion-related signals, such as walking, running, and even rapid leg bouncing. The Butterworth filter was chosen for its maximally flat frequency response in the passband. It does not introduce sharp transitions or excessive distortion, preserving the data integrity while filtering out noise.

To ensure consistency in the dataset and exclude non-protocol-related activity, only the last 20 minutes of accelerometry data for each subject were included in the analysis. For participants with less than 20 minutes of data, all their data were retained for analysis. This approach aimed to exclude any potential noise, such as preparation or calibration periods at the beginning of the recording, and focus the analysis on the structured activity protocol. Importantly, no specific activities were intentionally removed; instead, this step was taken to enhance the quality and relevance of the dataset.

\subsection{Classification Approaches}

\subsubsection{\bf Cut-Point Based Classification}. \label{sec:binary_method}

As a first approach to characterizing motion, the aim was to classify activity levels into a binary outcome: active or inactive. This distinction is a critical step in interpreting biomedical data such as ECG and heart rate. Activity levels influence the interpretation of heart rate variability, as the expected ranges for heart rate and ECG patterns differ based on whether a subject is at rest or in motion. Moreover, it is critical to understand if changes in HRV are endogenous (related to a change in physiology) or exogenous (due to a change in physical activity). Therefore, distinguishing between active and inactive states provides valuable metadata to inform clinical assessments.

To accomplish this, the vector magnitude (also known as the Euclidean norm) of the accelerometry data was computed. This calculation combined the filtered signals from the x, y, and z axes into a single value at each time point using the formula $\sqrt{x_{\mathrm{filtered}}^2 + y_{\mathrm{filtered}}^2 + z_{\mathrm{filtered}}^2}$. This provided a measure of the overall movement intensity, taking into account the contributions of all three axes of the accelerometry signal.

To classify activity levels, a rolling threshold approach was applied to the vector magnitude values. A rolling threshold involves evaluating the accelerometer signal magnitude over a moving time window. At each time point, the rolling median of the magnitude values is calculated within a specified 5-second window. This method helps smooth out short-term fluctuations and highlights trends over time, facilitating a more accurate classification of activity states. The 5-second window was selected based on its ability to capture the dynamic nature of human movements while maintaining a balance between temporal resolution and signal stability. Previous studies have shown that mid-sized windows size (from 5 to 7 seconds long) allow for capturing sufficient movement characteristics of an activity while also enabling the quick detection of transitions between activities, striking a balance between detailed analysis and real-time responsiveness \cite{twomey2018comprehensive, janidarmian2017comprehensive}. A rolling median was specifically chosen because it is more robust to outliers in the accelerometry signal, such as sudden spikes caused by device artifacts or environmental vibrations, ensuring a reliable signal for subsequent analysis and classification.

\subsubsection{\bf Multi-Class Classification.} \label{sec:multi_class_method}

For the second approach, a machine learning methodology was employed to classify multiple distinct activities based on accelerometry data. A 1-dimensional Convolutional Neural Network (CNN) was chosen due to its ability to effectively capture temporal patterns and hierarchical features within sequential data, such as accelerometry time series. CNNs have been widely used in time-series classification tasks, as they are capable of learning spatial and temporal correlations in sequential data, making them well-suited for detecting activity-related patterns \cite{hammerla2016deep, chen2021deep}. This approach aimed to recognize and differentiate between five activities: lying, sitting, standing, walking, and jogging.

As detailed in Section~\ref{sec:preprocess}, all accelerometry signals were first preprocessed using a Butterworth bandpass filter with cutoff frequencies of 0.05 Hz and 2 Hz to remove drift and high-frequency noise. To validate the robustness of this choice, a sensitivity analysis was performed by systematically varying the high-frequency cutoff to 5, 10, 15, and 20 Hz, as described in Section~\ref{sec:cutoff}. This analysis was conducted to ensure the filtering strategy preserved the most relevant human activity signals while minimizing irrelevant noise.

\begin{figure}[!htbp]
    \centering
    \includegraphics[width=\textwidth]{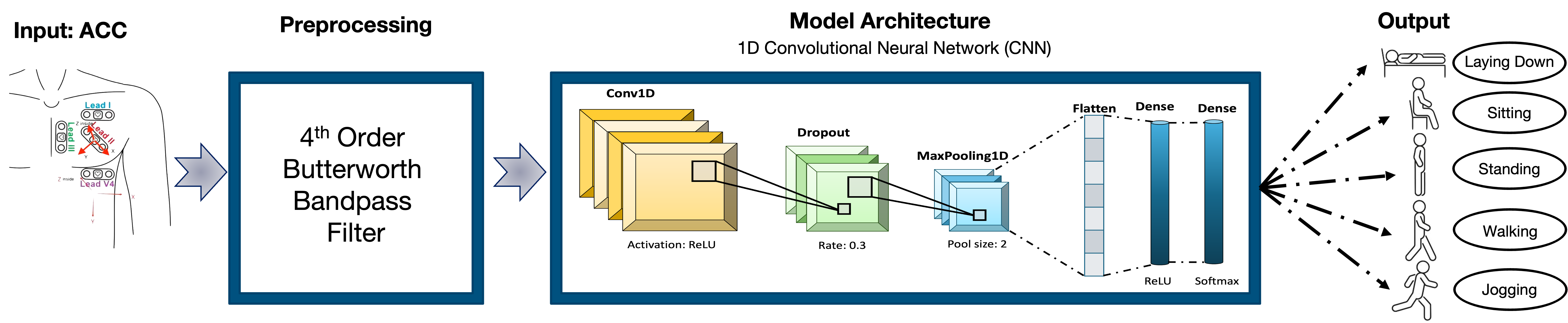}
    \caption{Architecture of the deep learning model for the activity classification.}
    \label{fig:cnn_architecture}
\end{figure}

The CNN architecture used in this study is illustrated in Figure~\ref{fig:cnn_architecture}, detailing the convolutional, pooling, and dense layers. The network was trained using data labeled with the five activity classes. \add{To account for the imbalanced distribution of activities, class-specific weighting was incorporated during training to ensure that underrepresented activity categories contributed proportionally to model optimization.} During training, the model learned to associate specific spatiotemporal patterns in the accelerometry signals with each class.

Model performance was evaluated using Leave-One-Out Cross-Validation (LOOCV). In this setup, data from one participant was held out for testing while the model was trained on the remaining participants. This process was repeated iteratively for all participants, ensuring that no individual’s data was seen during training on the fold where they were tested. This approach provides a rigorous evaluation of generalization performance in small datasets.

To quantify model performance, accuracy, precision, recall, and F1-score were computed using weighted averaging. \add{Weighted metrics reflect the prevalence of each activity class and provide an estimate of overall performance under naturally imbalanced activity distributions.}
Alternative methods, such as macro and micro averaging, were considered but found less appropriate given the highly imbalanced nature of the dataset (e.g., overrepresentation of sitting and underrepresentation of jogging). Weighted metrics therefore offer a more realistic estimate of model utility in clinical or real-world deployments where minority classes are equally important.

To evaluate how model performance is affected by different parameters, input window sizes (1s, 2s, 4s, 5s, 8s, 12s) and sampling rates (5 Hz, 10 Hz, 25 Hz, 50 Hz) were systematically varied. While a 5-second window and 50 Hz sampling rate were justified as the default based on prior literature and empirical optimization, exploring a broader range of parameter settings provides important insights into model robustness, especially under real-time or resource-constrained scenarios. These analyses support the development of adaptable systems that can maintain accuracy under varying computational or sensor limitations.

To further assess whether observed differences in performance were attributable to experimental design choices or participant-level variability, a linear mixed-effects model was applied. The outcome metric (e.g., F1-score) was modeled as a function of window size and sampling rate, which were treated as fixed effects due to their systematic manipulation across experiments. Subject ID was modeled as a random effect to account for inter-individual differences in activity performance or sensor placement. This modeling approach allowed us to disentangle the influence of these design parameters from subject-specific noise and better evaluate their generalizability. Performance scores from all window-sampling rate combinations were included as input to the model. Full results are reported in Section~\ref{sec:multi-class}.

\subsection{Resting Vitals}
This analysis builds upon the cut-point based classification method described in Section \ref{sec:binary_method} by evaluating the physiological relevance of the rolling 3D magnitude signal. Specifically, the analysis examined whether greater movement intensity, as captured by the 5-second rolling magnitude, is associated with deviations from each subject’s resting heart rate. This relationship supports the use of motion-derived features for contextualizing cardiovascular responses in ambulatory settings.

Each subject’s resting heart rate was estimated by computing the median heart rate during periods labeled as ‘lying’ or ‘sitting’ prior to the first recorded instance of walking. Heart rate deviation was then calculated as the difference between each observed heart rate value and the individual’s baseline. To quantify movement intensity, the rolling 3D magnitude measure derived from accelerometer data was used. Pearson correlation analysis was performed to evaluate the relationship between rolling magnitude and heart rate deviation, and a scatter plot was generated to visualize the trend. This approach provides insight into the potential of using physical activity features as proxies for physiological stress or exertion.


\section{Results}\label{sec:results}
\subsection{Cut-Point Based Classification Performance}



This section reports the performance of the cut-point based classification method in distinguishing active and inactive states using accelerometry data.

\add{A threshold of 0.07 was selected based on empirical distribution analysis of rolling 3D magnitude values obtained from pilot recordings collected prior to the main study. Histograms were generated for inactive states (lying, sitting, standing) and active states (walking, jogging), and candidate thresholds were evaluated by examining the degree of overlap between the two distributions. The value of 0.07 was chosen as the threshold that minimized the number of windows falling within the intersection of the active and inactive distributions, thereby reducing misclassification between the two groups. The selected threshold was subsequently spot-checked on a subset of the main study dataset to confirm that the separation generalized adequately before being applied to the full cohort.}


Subsequently, 80\% of the subjects from the main dataset (with no participant overlap between training and testing sets) were used to assess the performance of this threshold.


Model performance was assessed using a held-out validation set comprising 20\% of participants. The model achieved an accuracy of 0.87, a precision of 0.73, and a recall of 0.90. This highlights the model’s ability to accurately distinguish activity levels under typical ambulatory conditions.


\subsubsection{{\bf Distribution Analysis.}}

The distribution of 3D magnitude values was compared between inactive and active states in the training set. 

\begin{figure}[!htbp]
    \centering
    \includegraphics[width=\textwidth]{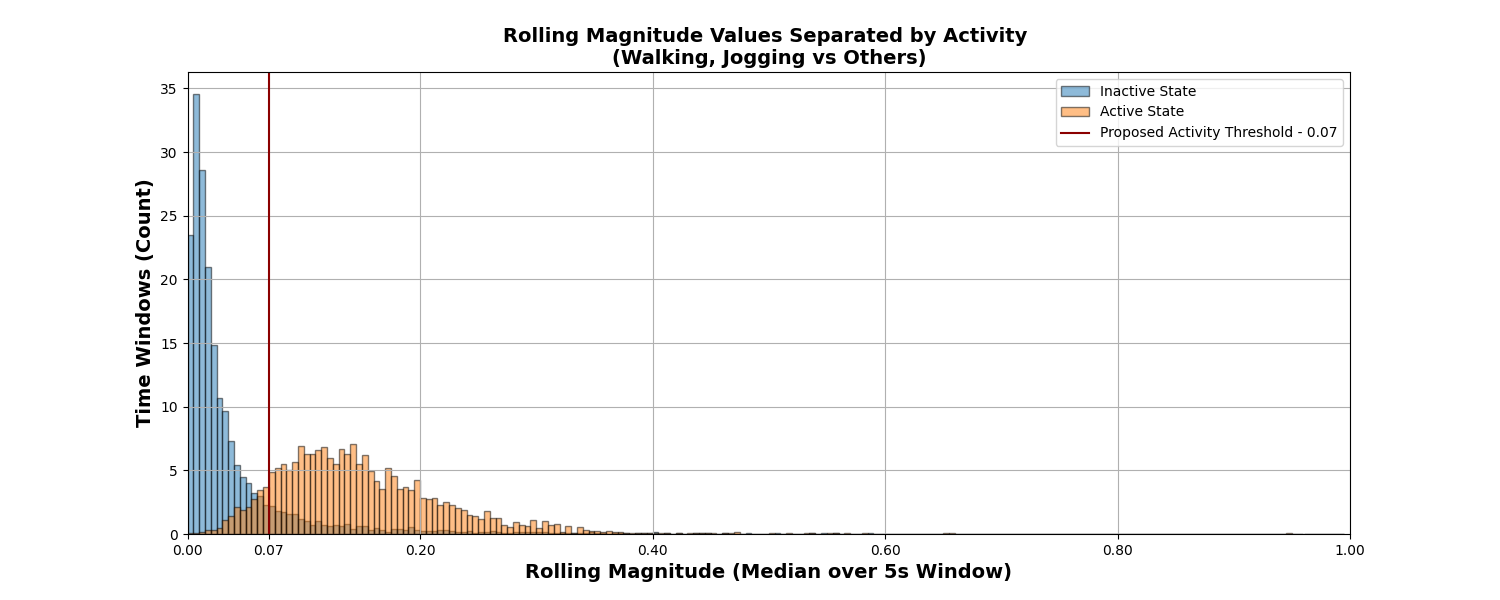}
    \caption{Histogram of Activity Count Values Separated by Activity.}
    \label{fig:active_vs_inactive}
\end{figure}

Figure \ref{fig:active_vs_inactive} presents the distribution of rolling magnitude values, computed as the rolling median of the 3D acceleration vector magnitude over a 5-second window. This representation allows for distinguishing inactive states (lying, sitting, standing) and active states (walking, jogging) using the proposed threshold of 0.07. Standing was categorized as an inactive state following prior literature \cite{kuster2020sitting}, which suggests that standing does not necessarily involve substantial movement and is often grouped with other inactive states in accelerometry-based studies.

\subsubsection{{\bf Confusion Matrix.}}

\begin{figure}[!htbp]
    \centering
    \includegraphics[width=0.5\textwidth]{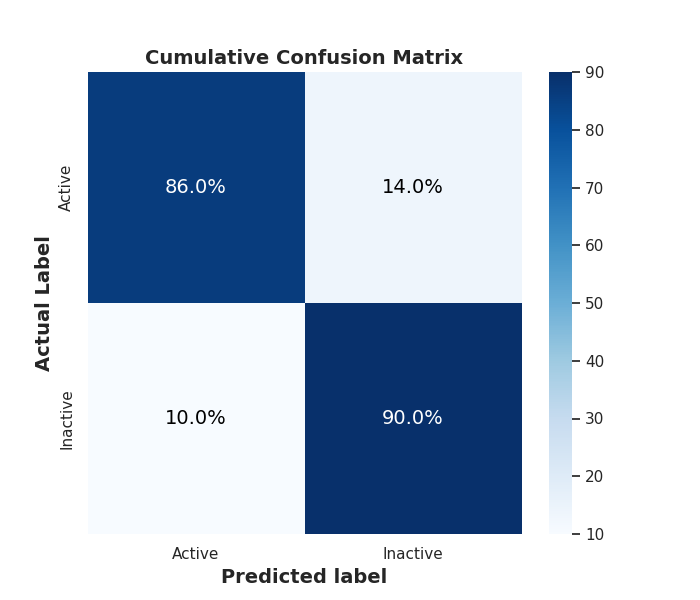}
    \caption{Confusion Matrix for Active vs. Inactive Labels.}
    \label{fig:conf}
\end{figure}

The confusion matrix, shown in Figure \ref{fig:conf}, provides a detailed breakdown of true positives, true negatives, false positives, and false negatives. This analysis offers more profound insight into the model's performance, highlighting areas of strength and potential misclassifications.

\subsubsection{{\bf Comparison with MAD-Based Cut-Point.}}

To assess how existing methods generalize to the current dataset, the Mean Amplitude Deviation (MAD)-based cut-point proposed by ~\cite{luckhurst2024classifying} was applied. Their study identified an optimal MAD threshold of 47.73\,mG (0.04773\,g) for classifying sedentary and ambulatory activity using the same VivaLink ECG Patch.

When this threshold was applied to the current dataset using a 5-second rolling MAD window, the method achieved an accuracy of 0.71. However, it failed to identify any active windows, resulting in a precision and recall of 0.00. This suggests the MAD cut-point is overly conservative in real-world ambulatory data and misclassifies most active states as inactive.

In contrast, the 3D magnitude-based threshold of 0.07 achieved substantially higher performance across all metrics, highlighting its robustness in noisier, more variable datasets. These results support the need for context-specific methods—including both the choice of activity metric and thresholding strategy—that are resilient to the variability and noise inherent in real-world ambulatory monitoring settings.

\subsection{Multi-Class Classification Performance } \label{sec:multi-class}

This section presents the results of the CNN-based multi-class activity classification using filtered triaxial accelerometry data. The model was trained to classify five distinct activity types—lying, sitting, standing, walking, and jogging—using LOOCV for evaluation. Performance was assessed under various preprocessing configurations to determine the impact of window size, sampling rate, and filtering parameters.


\subsubsection{\bf Baseline Model Performance.}

The baseline model was evaluated using a 5-second input window and a 50 Hz sampling rate. As summarized in Table~\ref{tab:performance}, accuracy, precision, recall, and F1-score were computed using weighted averaging. The model achieved a weighted accuracy of 77\%, precision of 83\%, recall of 77\%, and F1-score of 79\%. These metrics reflect the model’s ability to generalize across participants.

\begin{table}[!htbp]
\centering
\caption{Weighted performance metrics of the CNN classifier.}
\resizebox{0.35\textwidth}{!}{%
\begin{tabular}{|c|c|}
\hline
\textbf{Metric}       & \textbf{Weighted Score} \\ \hline
\textbf{Accuracy}     & 77\%                    \\ \hline
\textbf{Precision}    & 83\%                    \\ \hline
\textbf{Recall}       & 77\%                    \\ \hline
\textbf{F1 Score}     & 79\%                    \\ \hline
\end{tabular}%
}
\label{tab:performance}
\end{table}

\add{To provide further insight into class-specific performance, per-class F1-scores were computed cumulatively across all folds. Despite being underrepresented in the dataset, the jogging class achieved an F1-score of 79\%, indicating strong recognition capability for this minority activity. For completeness, the F1-scores for the remaining classes were: lying (55\%), sitting (82\%), standing (44\%), and walking (90\%). These results confirm that the model maintained meaningful discriminative performance across all activity categories and did not implicitly neglect rare classes.}

\subsubsection{\bf Distribution and Confusion Matrix.}

\begin{figure}[!htbp]
    \centering
    \begin{minipage}{0.48\textwidth}
        \centering
        \includegraphics[width=\textwidth]{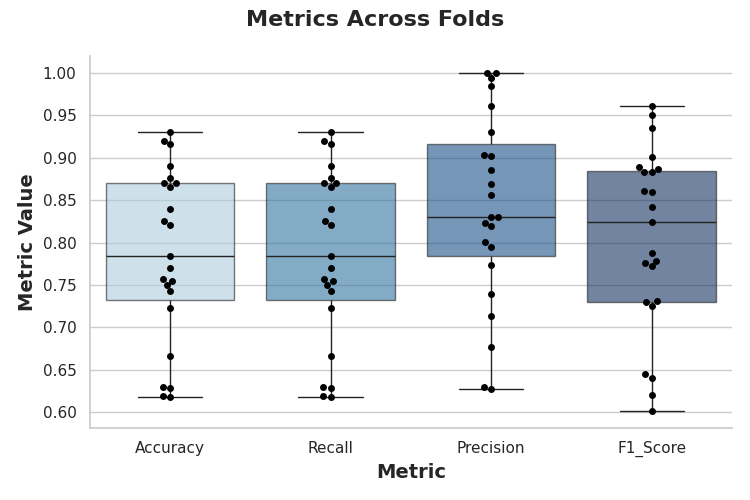}
        \caption{Box Plots of Different Metrics Across Different Folds for 5s Window Size.}
        \label{fig:boxPlot}
    \end{minipage}
    \hfill
    \begin{minipage}{0.48\textwidth}
        \centering
        \includegraphics[width=0.9\textwidth]{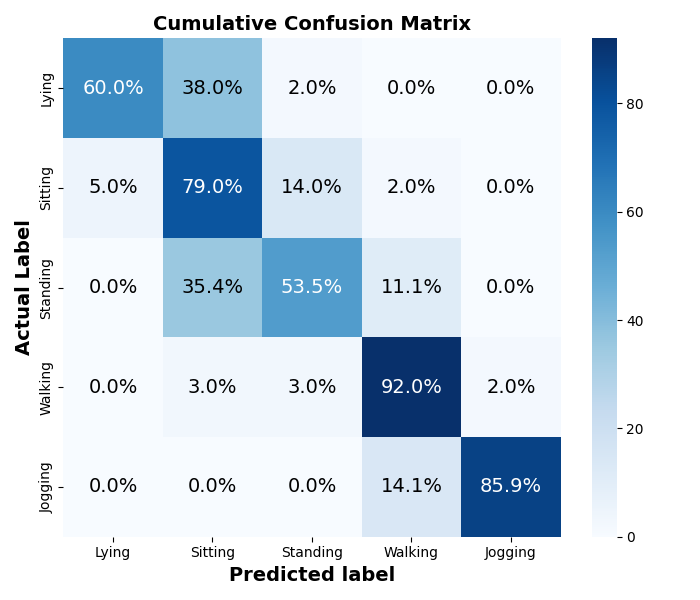}
        \caption{Confusion Matrix for 5s Window Size.}
        \label{fig:conf5s}
    \end{minipage}
\end{figure}

To assess consistency across participants, box plots of evaluation metrics are shown in Figure~\ref{fig:boxPlot}. The relatively narrow distributions confirm stable performance of the CNN across individuals.

A confusion matrix was also included to further illustrate the model’s classification performance. This matrix, presented in Figure~\ref{fig:conf5s}, highlights that the base model tends to misclassify somewhat similar activities. Notably, the model frequently confuses sitting and standing, as both postures exhibit similar vertical (z-axis) fluctuations, making them challenging to differentiate with a single accelerometer. In many instances, standing is misclassified as walking, particularly when the 5-second window includes minor walking movements or other motions that contaminate the standing data. However, the model does not confuse more distinct activities, such as walking and lying down, which indicates its ability to differentiate between activities with varying motion patterns.

\subsubsection{\bf Sensitivity of CNN Classification to Cutoff Frequencies.}\label{sec:cutoff}

\begin{figure}[!htbp]
\centering
\includegraphics[width=\textwidth]{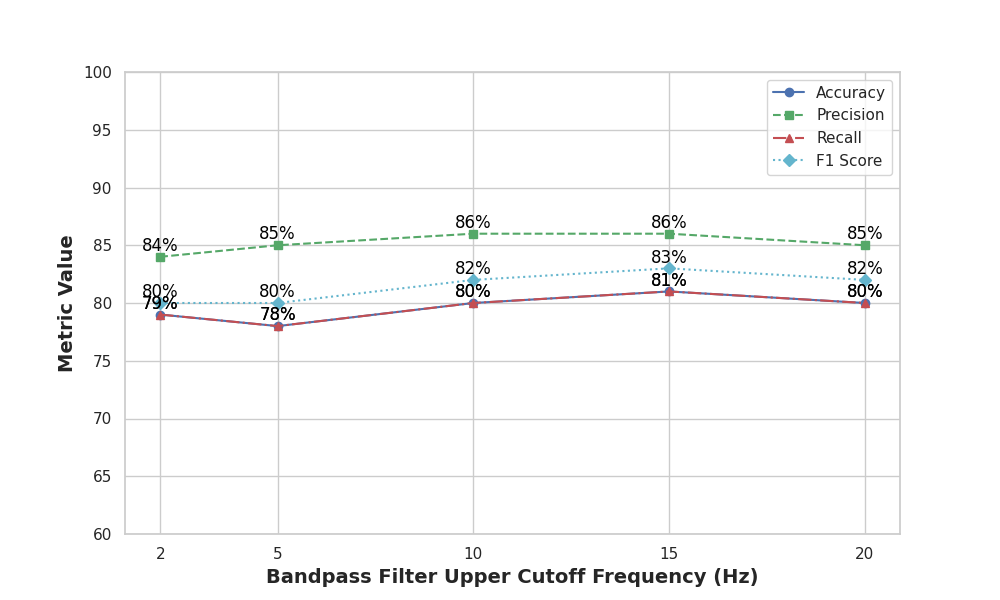}
\caption{Classification performance across different high-frequency cutoff values in the Butterworth filter.}
\label{fig:cutoff}
\end{figure}

To examine how the CNN classification performance responds to variations in the filtering step described in Section~\ref{sec:preprocess}, a sensitivity analysis was conducted by varying the high-frequency cutoff in the Butterworth filter (2, 5, 10, 15, and 20 Hz). As shown in Figure~\ref{fig:cutoff}, classification performance declined as the cutoff frequency increased, suggesting that frequencies above 2 Hz introduce noise without improving discriminatory information. These results support the original filter design and demonstrate that lower cutoff frequencies preserve relevant motion signals for activity classification.

This experiment was conducted using the multi-class CNN model to ensure that conclusions about filter design directly inform model performance.




\subsubsection{\bf Impact of Varying Input Size and Sampling Rate.}



In this section, the effect of varying the window size and sampling rate on the model’s performance was investigated. The tested window sizes were 1, 2, 4, 5, 8, and 12 seconds, and the sampling rates were 5 Hz, 10 Hz, 25 Hz, and 50 Hz. The performance metrics—accuracy, recall, precision, and F1 score—generally decreased with shorter window sizes and lower sampling rates, likely due to the limited temporal context available in smaller windows. This reduction in information adversely affected the model’s ability to accurately classify activities.

\begin{figure}[!htbp]
    \centering
    \includegraphics[width=\textwidth]{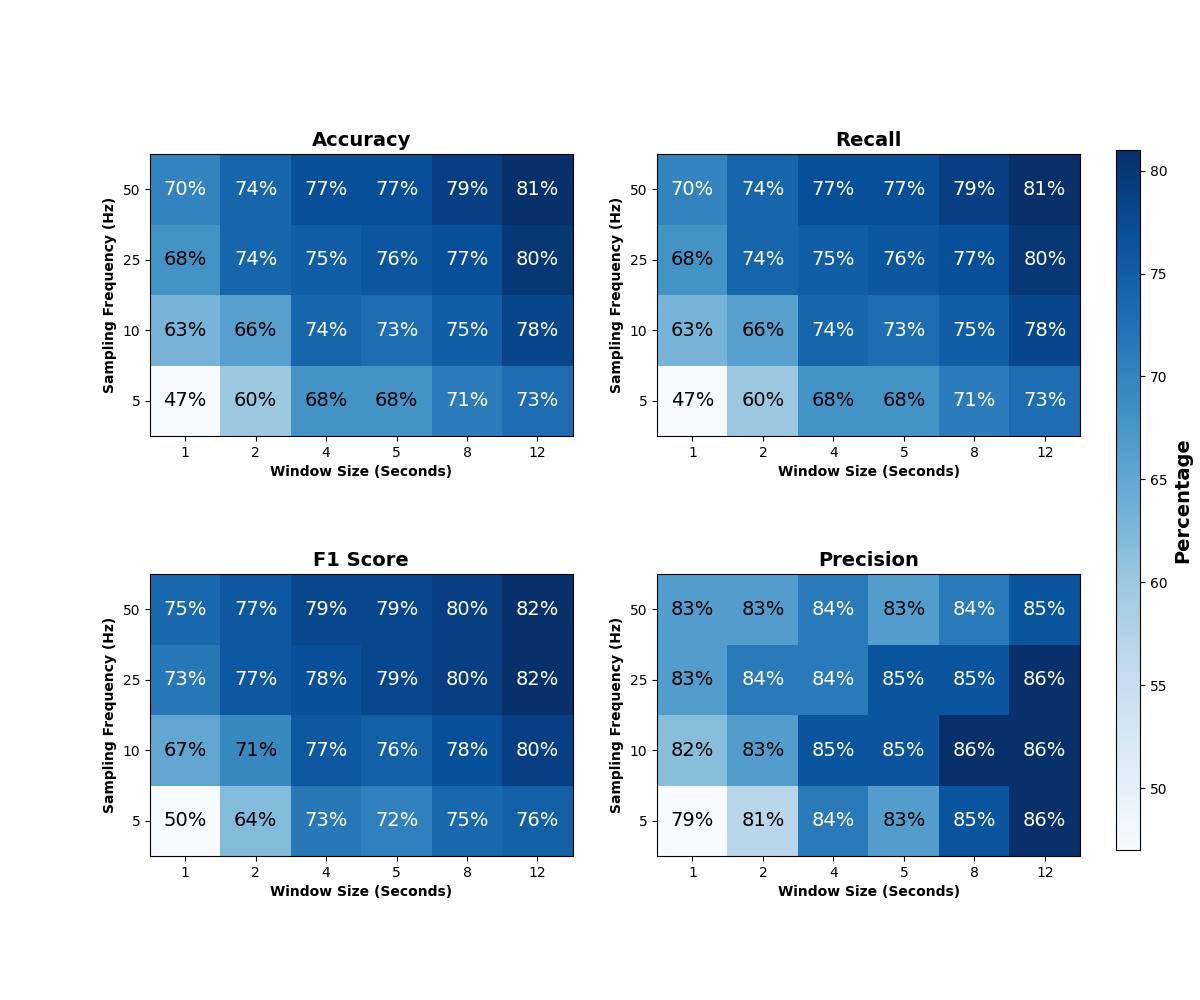}
    \caption{Heatmap of Different Metric vs. Window Size and Sampling Rate.}
    \label{fig:heatmap}
\end{figure}

To visualize these impacts, heatmaps were generated for each metric (Figure~\ref{fig:heatmap}), showing how different combinations of window size and sampling rate influence model performance. These plots provide insights into the trade-offs between temporal resolution and classification accuracy.



\subsubsection{\bf Mixed-Effect Model Analysis.}

This section presents the results of a mixed-effect model analysis to assess whether the observed performance differences are influenced by subject-specific factors or driven primarily by experimental parameters. The model quantified the impact of window size and sampling rate on classification metrics while controlling for inter-subject variability. Table~\ref{tab:mixed_model_results} summarizes the estimated coefficients and confidence intervals.




\begin{table}[!htbp]
\caption{Mixed Linear Model Regression Results}
\centering
\resizebox{\textwidth}{!}{%
\begin{tabular}{|c|c|c|c|c|c|c|}
\hline
\textbf{Predictor} & \textbf{Coef.} & \textbf{Std. Err.} & \textbf{z} & \textbf{$P > |z|$} & \textbf{[0.025, 0.975]} \\ \hline
Window Size 1 s vs. 12 s & \multicolumn{1}{r|}{-12.389} & \multicolumn{1}{r|}{0.594} & \multicolumn{1}{r|}{-20.853} & \multicolumn{1}{r|}{0.000} & \multicolumn{1}{r|}{[-13.553, -11.224]} \\ \hline
Window Size 2 s vs. 12 s & \multicolumn{1}{r|}{-7.264} & \multicolumn{1}{r|}{0.594} & \multicolumn{1}{r|}{-12.226} & \multicolumn{1}{r|}{0.000} & \multicolumn{1}{r|}{[-8.428, -6.099]} \\ \hline
Window Size 4 s vs. 12 s & \multicolumn{1}{r|}{-3.516} & \multicolumn{1}{r|}{0.594} & \multicolumn{1}{r|}{-5.919} & \multicolumn{1}{r|}{0.000} & \multicolumn{1}{r|}{[-4.681, -2.352]} \\ \hline
Window Size 5 s vs. 12 s & \multicolumn{1}{r|}{-3.120} & \multicolumn{1}{r|}{0.594} & \multicolumn{1}{r|}{-5.251} & \multicolumn{1}{r|}{0.000} & \multicolumn{1}{r|}{[-4.284, -1.955]} \\ \hline
Window Size 8 s vs. 12 s & \multicolumn{1}{r|}{-1.736} & \multicolumn{1}{r|}{0.594} & \multicolumn{1}{r|}{-2.923} & \multicolumn{1}{r|}{0.003} & \multicolumn{1}{r|}{[-2.901, -0.572]} \\ \hline
Sampling Rate 5 Hz vs. 50 Hz & \multicolumn{1}{r|}{-8.786} & \multicolumn{1}{r|}{0.485} & \multicolumn{1}{r|}{-18.113} & \multicolumn{1}{r|}{0.000} & \multicolumn{1}{r|}{[-9.737, -7.835]} \\ \hline
Sampling Rate 10 Hz vs. 50 Hz & \multicolumn{1}{r|}{-3.361} & \multicolumn{1}{r|}{0.485} & \multicolumn{1}{r|}{-6.928} & \multicolumn{1}{r|}{0.000} & \multicolumn{1}{r|}{[-4.314, -2.409]} \\ \hline
Sampling Rate 25 Hz vs. 50 Hz & \multicolumn{1}{r|}{-0.783} & \multicolumn{1}{r|}{0.485} & \multicolumn{1}{r|}{-1.613} & \multicolumn{1}{r|}{0.107} & \multicolumn{1}{r|}{[-1.733, 0.168]} \\ \hline
Accuracy vs. F1 Score & \multicolumn{1}{r|}{-3.284} & \multicolumn{1}{r|}{0.485} & \multicolumn{1}{r|}{-6.771} & \multicolumn{1}{r|}{0.000} & \multicolumn{1}{r|}{[-4.235, -2.334]} \\ \hline
Precision vs. F1 Score & \multicolumn{1}{r|}{8.333} & \multicolumn{1}{r|}{0.485} & \multicolumn{1}{r|}{18.333} & \multicolumn{1}{r|}{0.000} & \multicolumn{1}{r|}{[7.942, 8.724]} \\ \hline
Recall vs. F1 Score & \multicolumn{1}{r|}{-3.284} & \multicolumn{1}{r|}{0.485} & \multicolumn{1}{r|}{-6.771} & \multicolumn{1}{r|}{0.000} & \multicolumn{1}{r|}{[-4.235, -2.334]} \\ \hline
Group Var & \multicolumn{1}{r|}{70.490} & \multicolumn{1}{r|}{2.675}& \multicolumn{3}{c|}{} \\ \hline
\end{tabular}%
}
\label{tab:mixed_model_results}
\end{table}

The results indicate that reducing the window size from 12 seconds to shorter durations leads to significant decreases in classification performance. For example, 1-second windows resulted in the largest negative effect ($\beta = -12.389$, $p < 0.001$). Similarly, decreasing the sampling rate from 50 Hz to 5 Hz or 10 Hz substantially reduced performance, with the most pronounced effect observed at 5 Hz ($\beta = -8.786, p < 0.001$).

In contrast, changing the evaluation metric revealed that precision tends to score higher than F1, while both accuracy and recall tend to score lower, relative to F1-score. The random effect variance for subject ID (70.490) confirms the presence of inter-individual variability, though the fixed effects dominate the performance trends. These results reinforce that window size and sampling rate are primary drivers of model performance, beyond subject-specific differences.



\subsection{Resting Vitals Analysis}

In this section, the observed relationship between movement intensity—quantified by the rolling 3D magnitude—and heart rate deviation from resting baseline—is reported. \add{A Pearson correlation analysis revealed a statistically significant but modest positive association ($r = 0.29$, $p < 0.0001$) between physical activity intensity and heart rate deviation.} 

\begin{figure}[htbp]
    \centering
    \includegraphics[width=\textwidth]{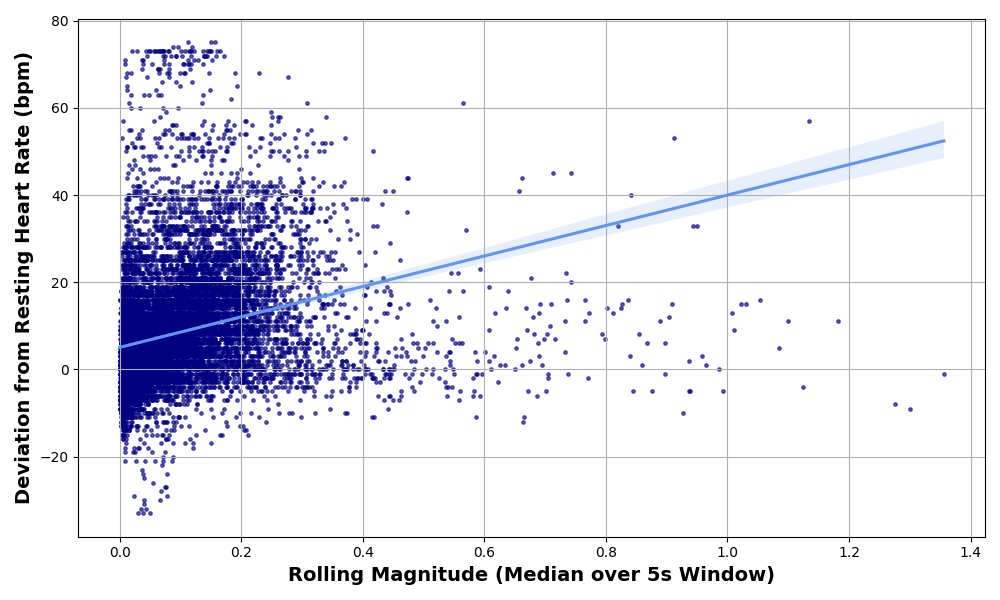}
    \caption{Relationship between Activity Count and Heart Rate Deviation.}
    \label{fig:3d_magnitude_vs_heart_rate}
\end{figure}

Figure~\ref{fig:3d_magnitude_vs_heart_rate} illustrates this relationship, where each point represents a 5-second window during an active period.

\section{Discussion}\label{sec:discussion}
The findings highlight the potential of 3D accelerometry for activity classification in ambulatory monitoring, with direct applications in healthcare diagnostics and patient monitoring. The ability to distinguish between different activity states, as demonstrated by the machine learning and signal processing approaches proposed in this study, is crucial for improving movement analysis in rehabilitation, chronic disease management, and elder care. For example, the results show that heart rate deviations correlate with physical activity, reinforcing the need to integrate motion tracking with physiological monitoring for more accurate health assessments. This finding aligns with previous studies that emphasize the value of contextualizing physiological data using motion sensors, such as in heart rate and oxygen saturation interpretation \cite{ivașcu2021activity, freeman2006autonomic}. This capability is particularly relevant for conditions like cardiovascular disease and congestive heart failure, where physical activity plays a critical role in disease management \cite{Greiwe2020wearable}. 

Although real-time classification was not implemented or tested in this study, all analyses were conducted retrospectively. Nevertheless, the methods demonstrated here provide a foundation for future development of real-time systems, as they use a 5 s sliding window with low per-update computational cost for the rolling-magnitude method ($\sim O(N)$ operations) and millisecond-scale inference time for the CNN framework. These findings suggest that accelerometry-based models could potentially be incorporated into continuous monitoring applications in clinical or home settings. Such integration is especially pertinent given the growing reliance on wearable technologies in remote and ambulatory care environments \cite{jat2022smart, albahri2019based, stevens2024feasibility}.
It is important to note that the model was not trained on transitional periods and is therefore rather unstable when presented with transitional data. The rapid fluctuations are typical of any hysteresis at a margin of transition. It is therefore recommended that periods of rapid fluctuations in the model output are labelled as ‘transition’ or 'unknown’.


Building on these findings, the combined use of 3D accelerometry data with signal processing and machine learning techniques has the potential to enhance the ability to manage variability in ambulatory settings caused by patient movement. The signal processing approach provides a computationally efficient method for distinguishing general activity levels, while the machine learning model, particularly the CNN, enables more detailed and precise activity classification, allowing for a more nuanced understanding of patient behavior. This supports prior work showing CNNs are well-suited for time-series accelerometry data in HAR applications \cite{shi2023novel, yuan2022interpretable}.

\add{A secondary finding of this study is that heart rate deviations demonstrated a modest correlation with physical activity intensity (r = 0.29), suggesting that motion contributes to, but does not fully explain, variability in heart rate responses. The resting vitals analysis underscores the importance of contextualizing physiological measurements within activity states. By establishing a subject’s resting heart rate based on periods of inactivity (lying or sitting), deviations in heart rate due to movement were quantified. This approach supports a more accurate interpretation of cardiovascular responses in ambulatory settings. Without accounting for activity levels, heart rate monitoring may misinterpret physiological stress, potentially leading to false alarms or misinterpretation of underlying conditions. For example, an increase in heart rate could be incorrectly attributed to stress or pathology when it may simply reflect a transition from rest to movement. These findings support the integration of motion tracking with heart rate monitoring to enhance the clinical utility of wearable health devices. \cite{pawar2007impact}}



While this study does not present a complete biomedical monitoring system, it demonstrates the feasibility of using accelerometry-based CNN models for accurate multi-class activity classification. Unlike device-specific pipelines that rely on idealized datasets or multi-sensor inputs, the studied approach was developed using noisy, imbalanced data collected under realistic conditions. This design choice directly addresses common limitations of HAR systems outlined in prior studies, including challenges with data quality, noise, and generalizability \cite{taffoni2018wearable, arshad2022human, ren2024clinical}.


Despite these promising findings, practical challenges remain in balancing model complexity with computational efficiency. The choice of window size and sampling rate significantly influences model performance, with longer windows and higher sampling rates generally yielding higher classification accuracy. However, these improvements come at the cost of increased computational demands, which may limit feasibility for real-time or resource-constrained applications. In this study, the observed performance gains with larger windows were likely driven by long stretches of homogeneous activity, which are easier to classify. In operational settings, however, longer windows may also capture transitions between activities—such as standing and walking—leading to ambiguity and potential misclassification. Future efforts should aim to optimize this trade-off, ensuring both accuracy and efficiency in more complex, real-world conditions.

One potential solution to this issue is to assign the activity label based on the most recent portion of the window. This approach assumes that the last segment of the window reflects the most relevant context for prospective applications or downstream decision-making. Since transitions often involve mixed activity patterns, focusing on the final portion of the window may help reduce misclassification due to overlapping behaviors, while still preserving the performance benefits of longer input sequences.

Despite the strengths of this study, several limitations must be acknowledged. \add{The study relied on a single chest-mounted accelerometer, which limits the ability to reliably differentiate between certain postures, particularly sitting and standing. From a biomechanical perspective, these postures exhibit similar trunk orientation and low acceleration variance when measured at the chest, explaining the observed confusion between these classes. In many ambulatory monitoring contexts, this misclassification may have limited clinical consequences, as both sitting and standing represent low-intensity states compared to walking or jogging. However, in applications where precise posture discrimination is clinically relevant—such as fall-risk assessment, orthostatic intolerance monitoring, or mobility rehabilitation—additional sensors such as thigh-mounted accelerometers or orientation-based features may be required to improve posture resolution \cite{wang2017review}.} Expanding the dataset to encompass a broader range of populations and real-world scenarios will also improve generalizability and clinical applicability.

\add{A further consideration relates to natural fidgeting behaviors during nominally static activities such as sitting and standing. These subtle movements were not assigned independent labels but were incorporated into the primary activity categories. While this design choice increases intra-class variability and may introduce ambiguity between activity classes, it reflects realistic ambulatory conditions in which individuals rarely remain perfectly motionless. Consequently, the model was trained to tolerate small postural adjustments rather than relying on artificially clean signals. Nevertheless, such fidgeting-related motion may partially explain residual confusion between sitting and standing, and future work could explore explicit modeling of micro-movements or posture sub-states to further improve discrimination.}

\add{Additionally, the cohort consisted exclusively of healthy adult volunteers rather than clinical patients. Therefore, while the findings demonstrate feasibility in ambulatory conditions, external validation in hospitalized or chronically ill populations is necessary prior to clinical deployment.}

\add{A further limitation relates to variability in recording duration across participants ($26.33\pm 21.36$ minutes). Some participants were unable to complete the full recording protocol due to discomfort, resulting in shorter time-series data. Although Leave-One-Out Cross-Validation reduces subject-specific overfitting, unequal recording lengths may introduce subtle bias in class distribution or temporal representation. Future studies should investigate the impact of recording duration on model performance and evaluate robustness under more standardized and balanced data-collection conditions.}

\add{A separate methodological consideration involves the treatment of ambiguous activity periods. In this study, transition windows and interference-labeled segments were excluded from training and evaluation to ensure well-defined class boundaries. While excluding transition windows improved classification performance for stable activity states, it can be interpreted as inflating performance metrics relative to real-world continuous monitoring, where transitional periods are common and unavoidable. In operational settings, transitions may introduce ambiguity and reduce classification stability. Nevertheless, it is inherently impossible to categorize states that were not defined as explicit classes. In fact, identification of this `transition' state remains an open question, and classification of how an individual moves between states may indeed be the most informative part of the activity. For example, the speed at which we transition from sitting to standing and vice versa is a very useful measure of aging \cite{BALACHANDRAN2021111202}. Therefore, one could imagine using our classifier as a method for identifying the transition points when the classifier fails to provide a high probability for a single class, enabling segmentation of the start and end of a transition.}
Future work should explore strategies for detecting or modeling such transitional states directly, which could enhance the completeness and robustness of real-world activity recognition \cite{arshad2022human}.

Moreover, an important limitation of this study is the absence of collected demographic variables such as race and education level. While the study’s primary aim was to develop and validate activity recognition methods, which are not expected to be primarily influenced by race (for example), the size of the cohort and the lack of demographic data prevent assessment of model generalizability across such factors. Future studies should prioritize collecting comprehensive demographic information to enable evaluation of potential biases and to improve the equity and applicability of activity classification models.

All data, code, trained models, and supporting documentation used in this research have been made publicly available \cite{cliffordlab2025}.


\section{Conclusion}\label{sec:conclusion}
This study successfully evaluated two approaches for recognizing patient activity using 3D accelerometry sensor data to improve the utility of biomedical monitoring in ambulatory settings. The signal processing approach effectively distinguished between high and low activity levels, while the machine learning algorithm demonstrated strong performance in classifying five distinct activities: lying down, sitting, standing, walking, and jogging. By accurately identifying activity states, these methods have the potential to help contextualize physiological measurements, reduce motion-induced inaccuracies and support the interpretation of wearable sensor data in healthcare applications. Additionally, the states themselves may offer diagnostic insights, as factors like time spent in a specific state or the frequency of transitions between states could be explored in future work for links to conditions such as over-sedation, delirium, or sleep-related illnesses. The proliferation of low cost and low-energy accelerometer devices, combined with advanced activity recognition techniques, holds promise to provide critical context for biomedical parameters, potentially enabling more reliable health monitoring in dynamic healthcare environments.

\section*{Acknowledgments} 

This study was partially funded by LifeBell AI LLC. Dr. Clifford holds equity in LifeBell. LifeBell has an interest in the type of software technology being developed and evaluated in the research described in this paper. 
The terms of this arrangement have been reviewed and approved by Emory University in accordance with its conflict of interest policies. 
Dr. Clifford is also supported by the National Center for Advancing Translational Sciences of the National Institutes of Health under award number UL1TR002378 and the National Institute of Biomedical Imaging and Bioengineering (NIBIB) under NIH award number R01EB030362.
The content is solely the responsibility of the authors and does not necessarily represent the official views of the National Institutes of Health, LifeBell AI LLC, or the authors' current and past employers and funding bodies.
This study was approved by WCG IRB (Pr. No.: 20231639).

\clearpage

\bibliographystyle{dcu}
\bibliography{refs}

\end{document}